\documentclass[10pt,letterpaper,twocolumn]{article} 

\usepackage{ol2}
\usepackage{amsmath}
\newcommand{\PRYSO}{Pr$^{3+}$:Y$_2$SiO$_5$ }

\begin{document}

\twocolumn[ 

\title{Photon echo without a free induction decay in a double-$\Lambda$ system}

\author{Sarah E. Beavan$^{1,*}$, Patrick M. Ledingham$^2$, Jevon J. Longdell$^2$ and Matthew J. Sellars$^1$}

\address{
$^1$Laser Physics Centre, RSPE, Australian National University, Canberra, ACT 0200, Australia\\
$^2$Jack Dodd Centre for Photonics  and Ultra-Cold Atoms, Department of Physics, University of Otago, Dunedin, New Zealand\\
$^*$Corresponding author: sarah.beavan@anu.edu.au
}

\begin{abstract}
We have characterized a novel photon-echo pulse sequence for a double-$\Lambda$ type energy level system where the input and rephasing transitions are different to the applied $\pi$-pulses. We show that despite having imperfect $\pi$-pulses associated with large coherent emission due to free induction decay (FID), the noise added in the echo mode is only $0.2\pm{0.1}$ photons per shot, compared to $4\times10^{4}$ photons in the FID modes. Using this echo pulse sequence in the `rephased amplified spontaneous emission' (RASE) scheme \cite{Ledingham2010} will allow for generation of entangled photon pairs that are in different frequency, temporal, and potentially spatial modes to any bright driving fields. The coherence and efficiency properties of this sequence were characterized in a \PRYSO crystal.
\end{abstract}


 ] 

\noindent 

An on-demand source of triggerable, transform-limited single photons is an integral component in many proposals for scalable quantum information processing and communication.  One approach outlined in the seminal paper of Duan, Lukin, Cirac and Zoller (DLCZ) \cite{Duan2001}, is to use an ensemble of atoms. A single excitation, if it is shared across many atoms over a distance of many wavelengths, has a well defined spatial mode given by the direction of constructive interference of the constituent atomic wavefunctions. The DLCZ protocol uses stimulated Raman transitions in $\Lambda$-type energy level systems to generate entangled pairs of photons. 

An alternative ensemble-based regime for generating non-classically correlated photon pairs proposed recently by Ledingham \emph{et al}\cite{Ledingham2010} uses photon echo techniques \cite{Abella1966} to rephase coherence which was created when a spontaneous emission event occurred.  The process begins by inverting the population in a two level system from its ground state into the excited state. Then for a period of time, the ensuing spontaneous emission events are recorded by a detector in the far field. These detected events herald an entangled state of the ensemble since we have no `which-way' information about where exactly the photon came from. Due to inhomogeneous broadening of the transition across the ensemble, the atoms will dephase relatively quickly. These collective atomic states can be rephased by applying a $\pi$ pulse, after which the ensemble will evolve and emit photon(s) identical to the initial spontaneous event(s) in a time-symmetric fashion about the refocusing $\pi$ pulse.  

An advantage of this RASE method compared to the Raman scattering regimes is that the bright driving fields are off while the single-photon fields are being generated (or rephased). Also, inherited from the photon echo upon which it is based, RASE is naturally temporally multi-mode. The major experimental difficulty in demonstrating RASE is in applying a precise $\pi$-pulse to facilitate the rephasing. In practice most of the atoms throughout the ensemble will experience pulse areas greater than or less than $\pi$, (for instance due to a non-uniform spatial beam profile) which leads to a large residual ensemble coherence generated by the pulse itself.  The coherent emission during this FID is generated in the same spatial and frequency mode as the $\pi$-pulse, which also defines the direction for optimal rephasing of the amplified spontaneous emission. Thus we are faced with the challenge of detecting the single rephased photon amidst this bright FID emission.  DLCZ type protocols, while lacking any temporal separability between pump and single-photon fields, do have the advantage that the single-photons are at a different frequency, and also can be generated in a different spatial mode from the bright driving fields.

In this paper, we propose and demonstrate a pulse sequence applied to a double-$\Lambda$ system that combines the advantages of the RASE and DLCZ protocols; allowing for the temporally multi-mode generation of single-photon states in \emph{different} temporal, frequency, and spatial modes to the bright pump beams.  We report on a demonstration of this `4-level echo' (4LE) in a rare-earth ion doped crystal. Aside from the immediate benefit of being an easy-to-implement solid state system, such crystals boast long coherence times \cite{Longdell2005, Fraval2004}, and are accordingly a promising system for development of quantum devices\cite{Hedges2010}.

\begin{figure}
	\centering
	\includegraphics[width=0.9\columnwidth]{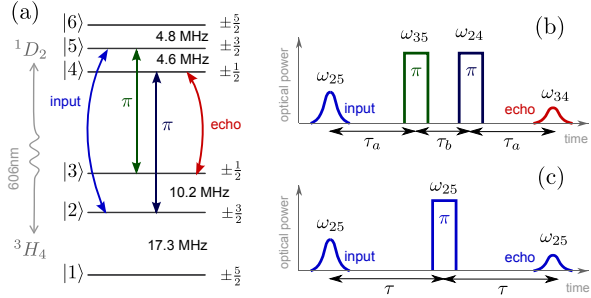}
\caption{\footnotesize(Color online) (a)Reduced energy level diagram for Pr$^{3+}$:Y$_2$SiO$_5$. (b)Pulse sequence for the 4LE, with transition frequencies marked. The total delay time between input pulse and echo is $2\tau_a+\tau_b$. (c)Pulse sequence for the standard 2-level echo used for comparison measurements of echo efficiency and decoherence. 
\label{fig:energylevels}}
\end{figure}

The sample used here was a $20\times4\times4$~mm 0.005\% \PRYSO crystal.  The transitions of interest occur between the $^{3}H_{4}$ and $^1D_2$ states of Pr in crystallographic site 1 \cite{Equall1995}.  These states each consist of three doubly-degenerate spin states with separations $\sim$10MHz (figure \ref{fig:energylevels}). The excited state lifetime ($T_1$) is roughly 160~$\mu$s, the coherence time ($T_2$) is 150~$\mu$s \cite{Equall1995}, and $T_1$ between ground state hyperfine levels is about 100~s. All optical transitions are weakly allowed, and the subset of levels $\left|2\right\rangle$, $\left|3\right\rangle$, $\left|4\right\rangle$ and $\left|5\right\rangle$ were chosen here as the double-$\Lambda$ system since transitions between these levels have similar strengths \cite{Nilsson2004}. The optical detection is performed using a balanced heterodyne system.


As depicted in figure \ref{fig:energylevels}(a) and (b), the 4LE sequence begins with an input pulse generating coherence on the $\left|2\right\rangle\rightarrow\left|5\right\rangle$ transition (frequency $\omega_{25}$), then after a delay time $\tau_a$ a $\pi$-pulse at $\omega_{35}$ transfers the coherence to a superposition between states $\left|2\right\rangle$ and $\left|3\right\rangle$. After a further delay $\tau_b$ a second $\pi$-pulse, at frequency $\omega_{24}$, is applied.  The combined effect of the two $\pi$-pulses is to conjugate the relative phase between the ions in exactly the same way as the single $\pi$-pulse in the 2LE, while also transferring the coherence to states $\left|3\right\rangle$ and $\left|4\right\rangle$. The ensemble rephases after time $\tau_a$, and an echo is generated. The key point to note is that the echo frequency ($\omega_{34}$) is different from each of the three other pulse frequencies.



A minimum of 4 energy levels is required to enable isolation of FID-related coherence from the echo transition; transferring the input coherence to a different frequency \emph{and} rephasing the ensemble requires a minimum of two rephasing pulses, and in tri-level echoes \cite{Mossberg1977}, there is inevitably a common energy level addressed by both bright pulses. Therefore the coherence generated by the first pulse is transferred to some extent by the second pulse onto the echo transition. 

The inhomogeneous broadening of the ${^3H_4}\rightarrow {^1D_2}$ transition is $\sim$5~GHz. This is much larger than the hyperfine splittings and thus any single frequency laser beam incident on the sample will simultaneously be resonant with different transitions for different ions throughout the crystal.  However to characterize the 4LE, we are only interested in a particular subset of ions that will be resonant with \textit{all} transitions shown in figure \ref{fig:energylevels}(a), and would prefer not to excite ions in alternative frequency classes that are irrelevant for the experiment and would affect the propagation of the pulses. So in order to characterize the 4LE, we first tailor the spectrum to pump all but one frequency subgroup of ions into a dark ground state.

The spectral holeburning process begins by iterating between 5 different frequency optical fields. Four of these fields are sweeping $\pm1$~MHz centred at frequencies $\omega_{25}$, $\omega_{35}$, $\omega_{24}$ and $\omega_{34}$, and the fifth field is at $\omega_{15}$. The purpose of this initial step, similarly to techniques described in \cite{Nilsson2004, Pryde2000}, is to pump away any population that is not in the desired frequency subgroup for the experiment. The next step is to create a spectral feature in state $\left|2\right\rangle$. This is achieved by iterating between the $\omega_{35}$ and $\omega_{34}$ sweeping fields (to ensure state $\left|3\right\rangle$ remains empty) and applying the $\omega_{15}$ field to pump a narrow spectral population back into state $\left|2\right\rangle$.  A typical feature-burning sequence takes $\sim$40~ms.

\begin{figure}
	\centering
	\includegraphics[width=0.9\columnwidth]{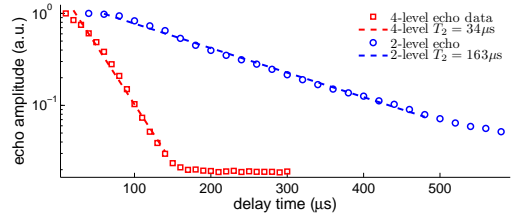}
\caption{\footnotesize(Color online) Photon echo amplitude with varying delay time. The x-axis represents the time between input pulse and the echo; 2$\tau_a$ for the 4-level sequence (here $\tau_b=0$) and 2$\tau$ for the 2-level sequence. Amplitude-FWHM of the $\pi$-pulses are 0.6$\mu$s and 1.0$\mu$s.
\label{fig:coherence_tau1}}
\end{figure}

Compared to the single-transition photon echo, the proposed 4-level sequence introduces additional inhomogeneity by utilizing 4 transitions.  The source of this inhomogeneity is on the hyperfine transitions, and is not rephased by the pulse sequence, so it is expected that the echo will decay faster than is the case for the 2-level system. As shown in figure \ref{fig:coherence_tau1} the 4LE decays with a time constant $T_2$ of 34~$\mu$s, nearly five times faster than the 2-level case. 

\begin{figure}
	\centering
	\includegraphics[width=0.9\columnwidth]{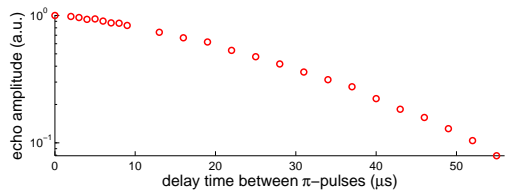}
\caption{\footnotesize(Color online) 4LE amplitude as a function of $\tau_b$, the delay time between the two $\pi$-pulses.  The loss of coherence is predominantly due to inhomogeneity on the RF transition, however the decay is not simply exponential, indicating a transition profile that isn't purely Lorentzian. For this data, $\tau_a=15~\mu$s.
\label{fig:coherence_tau2}}
\end{figure}

Long term storage of light in atomic media is achieved by transferring the coherence initially on an optical transition to the long-lived hyperfine ground states. This process is an inherent property of the 4LE sequence; the initial $\pi$-pulse transfers the coherence to the ground states, where it can be stored until the second $\pi$-pulse is applied to rephase on the optical transition. Figure \ref{fig:coherence_tau2} demonstrates the decay of the echo with increasing $\tau_b$.  Although the individual ions can remain in their superposition states for seconds, the ensemble coherence decays at a rate given by the inhomogeneity of the RF transition, or in this case a convolution of various transition inhomogeneities.  The decay trend in figure \ref{fig:coherence_tau2} is not consistent with a pure Lorentzian, nor a Gaussian lineshape, however the rate of decay indicates an inhomogeneous width on the order of 10~kHz. To take advantage of the long lifetime of the ground state coherences, it would be necessary to rephase the RF transition (similarly to \cite{Longdell2005}), then it becomes possible to extend the coherence time out to seconds \cite{Fraval2005, Fraval2004}.

\begin{figure}
	\centering
	\includegraphics[width=0.80\columnwidth]{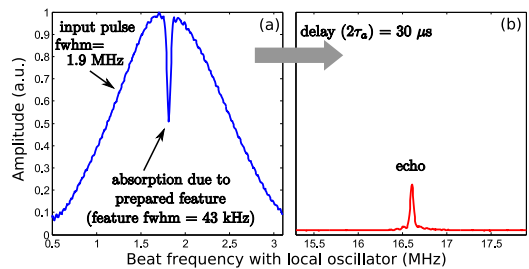}
\caption{\footnotesize(Color online) Example amplitude spectrum obtained for the 4LE efficiency measurements. (a)The input pulse; peak power is 5$\mu$W, or 0.3$\%$ of the $\pi$-pulse peak power. (b)The echo obtained after a total delay time (2$\tau_a$) of 30$\mu$s ($\tau_b=0$). The x-axis is the frequency of the signal after mixing with a local oscillator beam.  The frequency difference between input pulse and echo is 14.8MHz.
\label{fig:efficiencyeg}}
\end{figure}

The efficiency of the 4LE was determined by preparing a feature of 50\% absorption, then measuring the echo amplitude for a weak input beam (0.3\% of $\pi$-pulse peak intensity).  An example measurement is shown in figure \ref{fig:efficiencyeg}. The input pulse has a Gaussian profile and is spectrally much broader than the prepared feature. This pulse is mostly transmitted, allowing the optical depth and frequency width of the feature to be determined.  The echo amplitude normalized to the input pulse amplitude for the 2-level echo was measured as 41\% at a delay time of $\tau\approx0.55T_2$. The 4LE was 21$\%$ efficient at delay $\tau_a \approx T_2$. Extrapolating back to zero delay gives an efficiency of 63$\%\pm5\%$ and 57$\%\pm5\%$ for the 2 and 4-level echoes respectiviely, from which we conclude that there is very little difference in the rephasing efficiency of the two sequences\cite{EfficiencyEndnote}.

The primary motivation for the 4-level photon echo sequence is to rephase atomic coherence at a frequency that is otherwise completely dark. To check whether this is indeed the case, spectra were recorded for the 4LE without an input pulse. Averaging over 7500 shots showed that the number of photons added in the echo mode per shot is $0.2\pm0.1$, which is consistent with the level of amplified spontaneous emission expected \cite{ASEEndnote}.  At each of the $\pi$-pulse frequencies the added noise per shot is $\sim$40~000.  Thus in the echo mode, we have successfully avoided the emission associated with the $\pi$-pulses.

The sensitivity of the echo measurement using the 4LE sequence with frequency filtering is down to the spontaneous emission level, and therefore in the regime where the concept of rephasing amplified spontaneous emission can be experimentally tested. The RASE experiment begins with a population inversion. To create this in the current system the preparation could be modified such that ground states $\left|2\right\rangle$ and $\left|3\right\rangle$ remain empty, and a bright pulse at frequency $\omega_{15}$ is applied to populate excited state $\left|5\right\rangle$. A 4LE of the spontaneous emission at $\omega_{25}$ is then obtained at $\omega_{34}$.

In addition to frequency filtering, the fact that the 4LE includes two $\pi$-pulses offers the ability for spatial selectivity. Unlike the standard echo sequence where the input and rephasing pulses must be co-linear, the 4LE can satisfy momentum conservation without all beams being co-linear. 

Here we have shown that using a double-$\Lambda$ energy level structure to do a photon echo avoids the problem of distinguishing the echo emission from the FID that follows imperfect $\pi$-pulses.  By rephasing the coherence on an otherwise un-used transition, we can frequency filter to select only echo emission. Additionally this pulse sequence allows for off-axis phase matching and therefore spatial filtering.  The 4-level sequence has a similar efficiency to the standard 2-level photon echo, however the coherence time is shorter due to additional inhomogeneity introduced by using four transitions.  This 4LE will be extremely valuable for rephasing amplified spontaneous emission to generate non-classically correlated photon pairs.

SEB and MJS acknowledge the support of the Australian Research Council. PML and JJL were supported by the New Economy Research Fund of the New Zealand Foundation for Research Science and Technology.


\begin{thebibliography}{10}
\bibitem{ASEEndnote}
Spontaneous emission is an incoherent process resulting from the population inversion, and is therefore unavoidable, unlike the FID which is a coherent process resulting from imprecise $\pi$-pulses.
\end{thebibliography}
\end{document}